\documentclass[aps, prl, twocolumn]{revtex4-1}

\pdfoutput=1
\usepackage{mathtools}
\usepackage{amsfonts}
\usepackage{amssymb}
\usepackage{ulem}
\usepackage{dsfont}
\usepackage{array}
\usepackage{graphicx}
\usepackage{braket}

\newcommand{\bibs}{C:/Users/Ethan/Dropbox/References/BibFile}

\begin{document}
\title{Quantum Limits on Material Response Factors for Optimized Radiative Heat Transfer}
\author{Ethan L. Crowell}
\email[]{ecrowell4@wsu.edu}
\author{Mark G. Kuzyk}
\affiliation{Department of Physics and Astronomy, Washington State University, Pullman, Washington  99164-2814}
\begin{abstract}
Through quantum mechanical considerations, we optimize the material response factor $|\chi|^2/\text{Im}\left[\chi \right]$, which plays a pivotal role in the fundamental limits of near-field radiative heat transfer (RHT). A comparison of the limits obtained to experimental data for select materials shows that current materials fall several orders of magnitude short of the optimized values, suggesting the possibility of significant improvement in the rate of radiative heat transfer between two bodies. This work informs material design efforts that seek to optimize RHT, as well as provides insights into the quantum origins of RHT and the theory of fundamental limits.
\end{abstract}
\maketitle

\textit{Introduction} -- Radiative heat transfer (RHT) -- the process by which heat is transferred between two bodies via photons -- is an important example of light-matter interactions that impacts potential applications in thermophotovoltaics,\cite{lener14.01, karal16.01} nanoscale cooling,\cite{guha12.01} and thermal imaging.\cite{boudr97.01, jones12.01} To understand RHT, one begins by considering two bodies at temperatures $T_1$ and $T_2$. The thermal motion of the electrons in body 1 results in radiation, which can be absorbed by body 2. The heat transfer is simply the difference between the energy flux from body 1 to body 2, $\phi_{1\mapsto 2}$, and body 2 to body 1, $\phi_{2\mapsto 1}$, and is given by
\begin{align}
H_{1\mapsto 2} &= \phi_{1\mapsto 2} - \phi_{2\mapsto 1}\nonumber\\
&= \int_0^\infty \Phi(\omega)\left[\Pi(\omega, T_1) - \Pi(\omega, T_2)\right),\label{eqn - RHT definition}
\end{align}
where $\Pi(\omega, T)$ is the Planck distribution function and $\Phi(\omega)$ is a temperature independent energy flux.\cite{mulet02.01}

It is clear from Eqn.~\ref{eqn - RHT definition} that RHT will be optimal when materials support bound polaritons at frequencies coinciding with the peak in the Plank distribution function.\cite{moles20.01} Much research has been carried out to understand the limitations of enhancements of RHT.\cite{pendr99.01, biehs10.01, joula10.01, mille15.01}  Through a clever use of energy conservation and reciprocity, Miller \textit{et al}  discovered there exists a fundamental limit to RHT\cite{mille14.01,mille15.01,mille16.01} which they expressed as a limit on the spectral function $\Phi$ at a polariton resonance. For extended media separated by a distance $d$ \cite{mille15.01} this is
\begin{align}
\frac{\Phi(\omega)}{\Phi_{BB}} \leq \frac{1}{4(kd)^2}\zeta_1(\omega)\zeta_2(\omega),\label{eqn - MRF limit extended media}
\end{align}
where $\Phi_{BB}=\frac{k^2A}{4\pi^2}$ is the flux of a blackbody, and $\zeta_i$ is the material response factor of body $i$
\begin{align}
\zeta_i(\omega) = \frac{|\chi_i(\omega)|^2}{\text{Im}\left[\chi_i(\omega)\right]}.\label{eqn - MRF definition}
\end{align}
These limits are fundamental in the sense that they are independent of accidentals such as shape and size and only depend on the linear susceptibility.

The limit is expressed at a single frequency because it is assumed that the system is at a surface plasmon or phonon polariton resonance.  As a consequence, the amplified near field RHT has a narrow linewidth and one assumes that contributions away from the peak are negligible.

More recently, these limits have been extended to include multiple scatterers and finite size effects.\cite{venka20.01} This followed from an analysis of the singular value decomposition of the relevant response quantities, leading to the more strict limit
\begin{align}
\Phi &\leq \sum_i\left[\frac{1}{2\pi}\Theta\left(\zeta_A\zeta_Bg_i^2 -1\right)\right.\nonumber\\
&+\left.
\frac{2}{\pi}\frac{\zeta_A\zeta_Bg_i^2}{\left(1+\zeta_A\zeta_Bg_i^2\right)^2}\Theta\left(1-\zeta_A\zeta_Bg_i^2\right)\right]\label{eqn - MRF limit singular value},
\end{align}
where $g_i$ are the singular values of the vacuum Maxwell Green's function.

It is clear that the MRF is a key figure of merit for RHT. In this letter we show that the MRF has itself fundamental limits, which ultimately derive from the canonical commutation relations of quantum mechanics. This thereby leads to a more fundamental upper bound on RHT. We optimize radiative heat transfer at the molecular level, thus providing a guide for developing better materials that can then in turn be nano-structured for further improvements.

\textit{Limits on molecular polarizability} -- In parallel to finding fundamental limits to RHT between bulk systems, a considerable amount of work has been invested into understanding the fundamental limits of the molecular polarizability and hyperpolarizabilities \cite{shafe13.01, kuzyk00.01, kuzyk00.02}. The tensor elements of the polarizability are given by the sum over states (SOS) expressions\cite{boyd09.01, orr71.01}
\begin{align}
\alpha_{jk}(\omega) = \frac{e^2}{\hbar}\sideset{}{'}\sum_n&\left(\frac{x^{j}_{0n}x^{k}_{n0}}{(E_{n0}-i\Gamma_{n0}-\omega)}\right.\nonumber\\
&+\left.
\frac{x^{k}_{0n}x^{j}_{n0}}{(E_{n0}+i\Gamma_{n0}+\omega)}\right),\label{eqn - SOS alpha}
\end{align}
where the prime on the sum indicates that the $n=0$ term is excluded, $E_{n0} = E_n-E_0$ are the energy differences between eigenstates, $\omega$ is the frequency of the incident electric field, $\Gamma_{n0}$ is the phenomenological damping factor, and $x_{n0}$ is the (n,0) matrix element of the position operator.

Note that infinitely many states contribute to the polarizability in Eqn.~\ref{eqn - SOS alpha}. However, not all of the terms are independent. Using the canonical commutation relation
\begin{align}
\left[x, p\right] &= i\hbar,
\end{align}
it is easily shown that
\begin{align}\label{eq:HamiltonCommute}
\left[x,\left[x, H\right]\right] &= -\frac{\hbar^2}{m_e},
\end{align}
for a mechanical hamiltonian $H$ that may include electromagnetic interactions.\cite{watki12.01}  Taking matrix elements  between states $\Bra{p}$ and $\Ket{q}$ on both sides of Eqn.~\ref{eq:HamiltonCommute} and inserting unity ($\sum_n \Ket{n} \Bra{n}$), one obtains the Thomas-Reiche-Kuhn sum rules
\begin{align}
\sum_n x_{pn}x_{nq}\left(E_n - \frac{1}{2}\left(E_p + E_q\right)\right) = \frac{N_e\hbar^2}{2m}\delta_{pq}\label{eqn - sum rules}.
\end{align}
This system of equations relates the transition moments and energies to each other and acts as a constraint on the SOS expression for the polarizability, from which a maximum value for $\alpha$ can be determined.  For the static case, it is easily shown that the maximum polarizability is\cite{kuzyk13.01}
\begin{align}
\alpha_\text{max} &= \frac{e^2\hbar^2N_e}{m_eE_{10}^2}.\label{eqn - alpha max}
\end{align}

\textit{From molecular to bulk} -- We can make a connection between RHT and electron-photon interactions by relating the molecular polarizability $\alpha$ to the bulk susceptibility $\chi$ according to
\begin{align}
\chi^{(1)}_{ij} &= N\Braket{\alpha^*}_{ij}\label{eqn - susceptibility definition},
\end{align}
where $\alpha^*$ denotes the dressed polarizability, which accounts for local field effects within the material, $\Braket{\dots}$ denotes an ensemble average over molecular orientations, and $N$ is the molecular number density.

We consider the bulk to be made up of one-dimensional elements so only one element in the polarizability tensor is nonzero, namely $\alpha_{zz}$.  The more general tensor case is straightforward to treat in principle, but we adopt the more simple scenario for clarity in the presentation.  This is not to say that the tensor properties cannot be used for specific device requirements.\cite{kuzyk89.03} We will also restrict our considerations to amorphous solids, thereby assuming these 1D elements to be randomly oriented.  Again, this assumption is made for simplicity and can be relaxed without issue.  Averaging over all orientations yields
\begin{align}
\Braket{\alpha} &= \int d\Omega a_{iI}(\Omega)a_{jJ}(\Omega)\alpha^*_{IJ}\label{eqn - averaging}\\
 &= \frac{2}{3}\alpha^*_{zz},\label{eqn - averaged polarizability}
\end{align}
where $a_{iI}(\Omega)$ is the Euler rotation matrix \cite{case66.01}.

The dressed polarizability is related to the bare polarizability by the local field factor $L^{(1)}(\omega)$: \cite{jacks96.01, boyd09.01,kuzyk17.01}
\begin{align}
\alpha^*(\omega) &= L^{(1)}(\omega)\alpha(\omega),
\end{align}
where the local field factor is given by
\begin{align}
L^{(1)}(\omega) &= \frac{3}{3 - 4\pi N\alpha (\omega)}.\label{eqn - local field factor}
\end{align}

Combining Eqns. \ref{eqn - susceptibility definition}-\ref{eqn - local field factor}, we can write the bulk susceptibility as
\begin{align}
\chi^{(1)}(\omega)&= \frac{2N\alpha(\omega)}{3 - 4\pi N\alpha(\omega)}.\label{eqn - susceptibility}
\end{align}
Then, it is straightforward to show that
\begin{align}
\zeta &= \frac{2}{3}N\frac{|\alpha(\omega)|^2}{\text{Im}\left[\alpha(\omega)\right]}.\label{eqn - molecule to bulk MRF}
\end{align}
Eqn.~\ref{eqn - molecule to bulk MRF} connects the material response factor to the quantum mechanical response of the molecular constituents. Indeed, it is apparent from Eqn.~\ref{eqn - molecule to bulk MRF} that the MRF only depends on the optical properties of the individual molecules and the number density of molecules making up the material, along with a factor of $2/3$ resulting from random orientations. Local field enhancements, which often play a pivotal role in optical interactions, play no role in radiative heat transfer between amorphous solids.

\textit{Optimizing the MRF} -- In order to make the optimzation problem tractable, we will adopt an essential state model. Namely, we assume only three states contribute to Eqn.~\ref{eqn - SOS alpha} and Eqn.~\ref{eqn - sum rules}. This is referred to as the three-level model (TLM) and gives the truncated SOS expression for the polarizability
\begin{align}
\alpha^\text{3L}&(\omega) = e^2 \nonumber\\
& \times \left[|x_{10}|^2\left(\frac{1}{E_{10}-i\Gamma_{10}-\hbar\omega}
+ \frac{1}{E_{10}+i\Gamma_{10}+\hbar\omega}\right)\right.\nonumber\\
&+\left.
|x_{20}|^2\left(\frac{1}{E_{20}-i\Gamma_{20}-\hbar\omega}
+ \frac{1}{E_{20}+i\Gamma_{20}+\hbar\omega}\right)\right].\label{eqn - SOS alpha truncated}
\end{align}

In order for the MRF to be defined, we need a reasonable approximation for the phenomenological damping factor $\Gamma_{n0}$. The minimum damping allowed by quantum mechanics is half the natural linewidth \cite{schif68.01, kuzyk06.03}, which is given by
\begin{align}
\Gamma_{n0} &= \frac{2}{3}e^2\left(\frac{\Omega_{n0}}{c}\right)^3|x_{n0}|^2.\label{eqn - damping}
\end{align}
Eqn.~\ref{eqn - damping} provides the best-case value for the MRF. As such, we use it below.

It is convenient to introduce several parameters that appear in the theory of fundamental limits in nonlinear optics for our use here.\cite{kuzyk13.01} We have already seen the maximum polarizability in Eqn.~\ref{eqn - alpha max}. The TRK sum rules also define the maximum transition moment
\begin{align}
x_\text{max}^2 &= \frac{N_e\hbar^2}{2m_eE_{10}}\label{eqn - x max}
\end{align}
and the scale invariant energy and moment parameters
\begin{align}
E = \frac{E_{10}}{E_{20}} \hspace{1em} \mbox{ and } \hspace{1em}
X = \frac{x_{10}}{x_\text{max}}.
\end{align}
The energy parameter $E$ quantifies the energy level spacing. For example, $E=1$ for degenerate excited states.  The two level model, on the other hand, results when $E_2 \rightarrow \infty$ so that $E=0$. We also have $E=0$ when the ground state is doubly degenerate.

We can now work to reduce the number of free parameters in the Eqn.~\ref{eqn - SOS alpha truncated}. The $\left(p,q\right)=\left(0,0\right)$ sum rule  gives
\begin{align}
|x_{02}|^2 &= E\left(x^2_\text{max} - |x_{01}|^2\right),
\end{align}
which allows us to write the three level polarizability as
\begin{align}
&\alpha^\text{3L} = \frac{\alpha_\text{max}}{2}\left[X^2\left(\frac{1}{1 - i\gamma_{10} - \tilde{\omega}} + \frac{1}{1 + i\gamma_{10} + \tilde{\omega}}\right)\right.\nonumber\\
&+\left.
E(1-X^2)\left(\frac{1}{E^{-1} - i\gamma_{20} - \tilde{\omega}} + \frac{1}{E^{-1} + i\gamma_{20} + \tilde{\omega}}\right)
\right],
\end{align}
where $\tilde{\omega}= \hbar\omega/E_{10}$,
\begin{align}
\gamma_{10} &= \frac{\Gamma_{10}}{E_{10}} = \frac{N_e\alpha_\text{FS}}{6}X^2\frac{E_{10}}{m_ec^2},\\
\gamma_{20} &= \frac{\Gamma_{20}}{E_{10}} = \frac{N_e\alpha_\text{FS}}{6E^2}(1-X^2)\frac{E_{10}}{m_ec^2},
\end{align}
and $\alpha_\text{FS}$ is the fine structure constant. The system can thus be described by the dimensionless parameters $E$ and $X$, as well as the parameters $E_{10}$ and $\omega$. Note that although it may seem more natural to describe the frequency in units of $E_{10}$, thereby consolidating the two into a single dimensionless parameter, the natural linewidths cannot be simplified in this manner; the rest energy of the electron defines an additional energy scale. In effect we have four dimensionless quantities: $E, X, \tilde{\omega},$ and $E_{10}/mc^2,$.

Our approach to optimizing the MRF will be to fix values for $E_{10}$ and $\omega$, then find the values for $E$ and $X$ that maximize the MRF.

\begin{figure}[htp]
\centering
\includegraphics{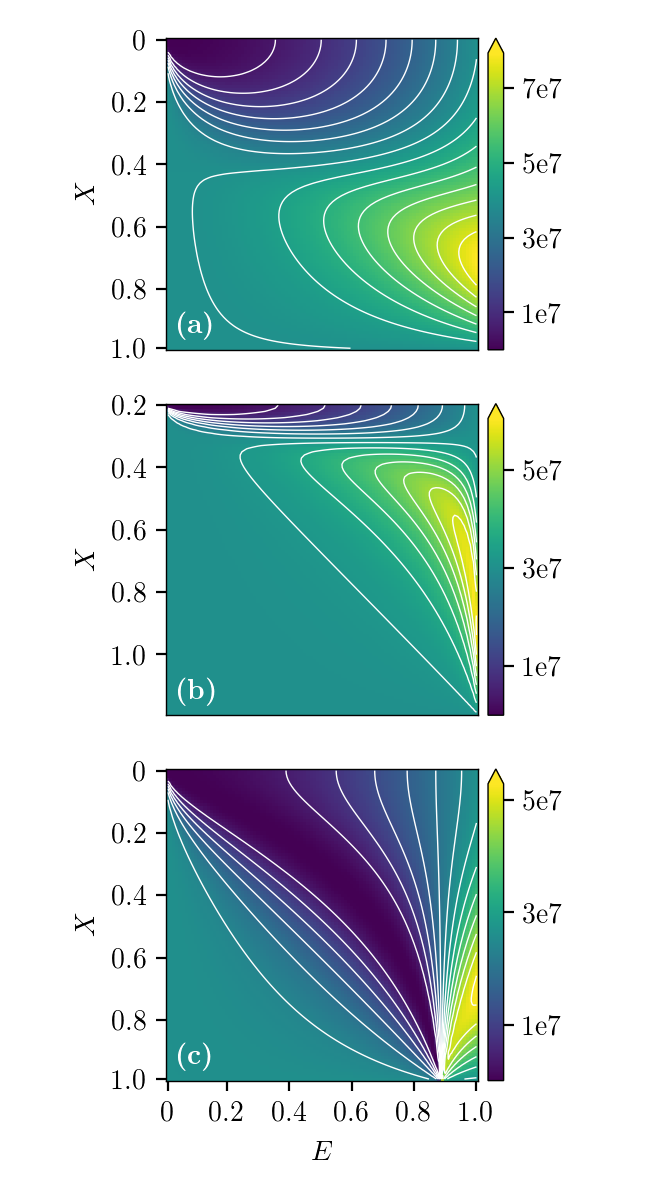}
\caption{Three level material response factor for $E_{10}=3.14$ eV and $N = 8.92\times 10^{-3} \, a_0^{-3}$, which are chosen to coincide with the corresponding experimental values for Aluminium. Plots are shown for (a) $\tilde{\omega} = 0.75E_{10}/\hbar$, (b) $\tilde{\omega} =  0.99E_{10}/\hbar$, and (c) $ \tilde{\omega} = 1.1E_{10}/\hbar$. Although the features change with frequency, the peak MRF remains fixed at $E=1.0$ and $X=0.71$.}
\label{fig - aluminium MRF}
\end{figure}
\textit{Results} -- The results of a raster scan through allowed values of $E$ and $X$ are shown in Fig.~\ref{fig - aluminium MRF} for select frequencies. Note that since $\alpha$ is symmetric in $X$, we restrict the view to $0\leq X\leq 1$. We choose $E_{10} = 3.14$ eV to coincide with the first electric-dipole transition in atomic Aluminium \cite{eriks63.01}, and $N = 8.92\times 10^{-3} \, a_0^{-3}$, the density of Aluminium in units of the bohr radius $a_0$. One is able to clearly identify regions in parameter space that result in a large or small MRF. The values corresponding to a maximum MRF are $E=1.0$ and $X=\pm 0.71$ for all frequencies.

\begin{figure}[htp]
\centering
\includegraphics{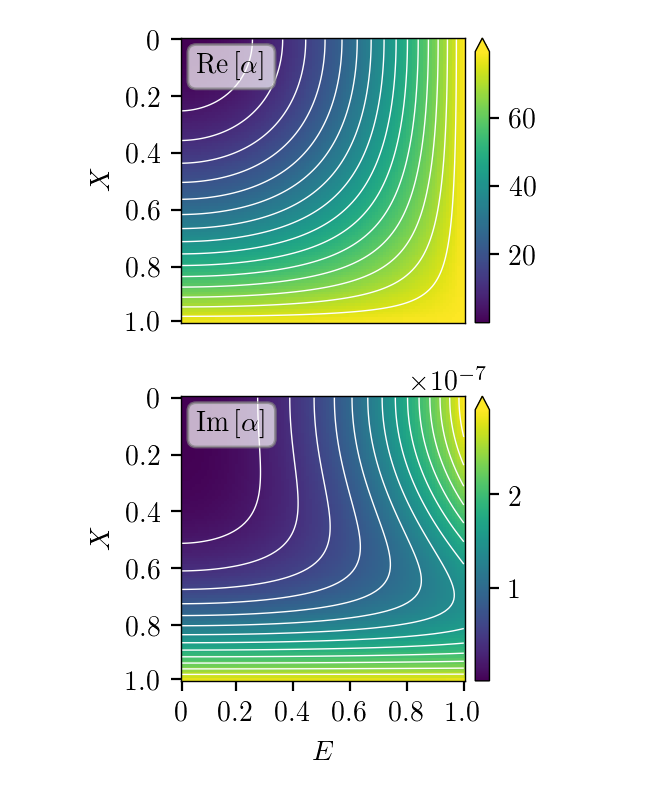}
\caption{(a) Real and (b) imaginary parts of the three level molecular polarizability for $E_{10}=3.14$ eV, $N = 8.92\times 10^{-3} \, a_0^{-3}$ and $\tilde{\omega} = 0.232$.  The real part is maximized along the bottom and right boundaries and absorption is minimum at $X=\pm 0.71$ when restricted to these boundaries. Thus, $E=1$ and $X=\pm 0.71$ represents the optimum trade-off between large response and low absorption.}
\label{fig - aluminium alpha real imag}
\end{figure}

Fig.~\ref{fig - aluminium alpha real imag} shows the real and imaginary parts, respectively, of the molecular polarizability over the aforementioned parameter space.  Linear scattering ($\text{Re}\left[\alpha\right]$) is maximum along the two boundaries defined by $X=\pm 1$ and $E=1$, respectively, while absorption ($\text{Im}\left[\alpha\right]$) is a minimum at $X=\pm 0.71$ if constrained to these boundaries.  Thus the requirement of a large response to an optical field must be balanced against a small absorption cross-section;  the former obtained when the two excited states are degenerate and the latter through tuning the spatial configuration of the wavefunctions (i.e. $X$).  A large transition moment between the ground and first excited state enhances the required light-matter interactions, but also increases absorption.

The TLA gives an approximation to the optimized MRF because truncation of states leads to contradictions in the TRK sum rules.\cite{kuzyk15.01} To provide an estimate of the disparity, we calculated the MRF using many states ($N=21$) in Eqn.~\ref{eqn - SOS alpha}, where the sum rules are enforced for $1\leq p,q\leq 7$. Details on the protocol that we use at select frequencies can be found in the work of Lytel \textit{et al} \cite{lytel17.01}. The largest MRF obtained with this method is approximately four times larger than the three-level limits. However, the optimal parameters lead to the same conclusions regarding material design; namely, degeneracies among the excited state are desirable for enhancing linear scattering, while tuning the transition moments between states leads to a decrease in absorption. This suggests that although the TLM doesn't provide an exact upper bound, it none-the-less provides a fundamental measure of the quality of the material response factor and provides key target parameters for the design of optimal materials for RHT.

We repeated the optimization procedure for characteristic energies $E_{10} = 4.63~\text{eV}$, $3.66~\text{eV}$, and $4.93~\text{eV}$, which coincide with the first electric-dipole transition of Gold \cite{ehrha71.01}, Silver \cite{picke01.01}, and Silicon \cite{radzi65.01}, respectively.  We also tried densities corresponding to silver ($8.68\times 10^{-3} \, a_0^{-3}$ ), gold ($8.75\times 10^{-3} \, a_0^{-3}$), and amorphous silicon ($7.40\times 10^{-3} \, a_0^{-3}$). Once again, for every combination of $E_{10}$ and number density $N$, we maximized the MRF with respect to $E$ and $X$. We found that the maximum value of the MRF depends on density and $E_{10}$, varying within the range $10^7-10^8 \, a_0^3$. Rather surprisingly, the optimal parameters $E$ and $X$ displayed no such dependence: the MRF was optimized with $E=1$ and $X=\pm 0.71$ for all densities and first excited state energies.

\begin{figure}[htp]
\centering
\includegraphics{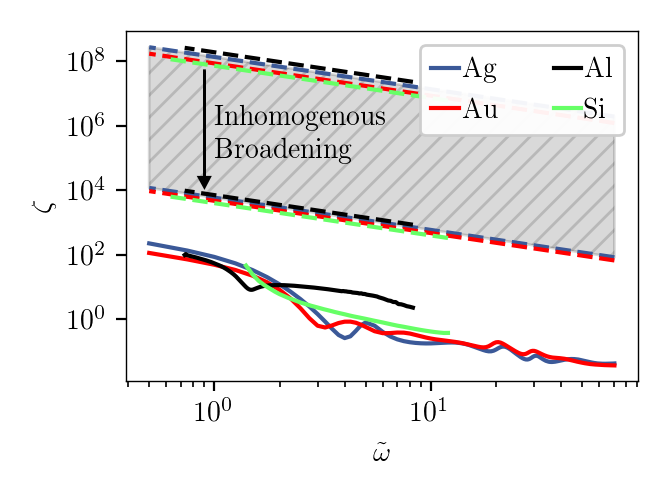}
\caption{Material response factors of several materials (solid lines) and the corresponding optimized three-level material response factors. The larger limits are those obtained by only including the natural linewidth, while the lower limit is obtained by including inhomogeneous broadening.}
\label{fig - MRF summary}
\end{figure}

\textit{Comparison to Experiment} -- Next, we study how actual materials compare to these optimized values. Tabulated complex permittivities for Silver \cite{werne09.01}, Gold \cite{werne09.01}, Aluminium \cite{mcpea15.01}, and Silicon \cite{pierc72.01} were used to calculate the polarizability of each material. From the polarizability, we used Eqn.~\ref{eqn - molecule to bulk MRF} to determine the MRF. These are shown in Fig.~\ref{fig - MRF summary} (solid lines) along with the optimized MRF from the TLM (upper set of dashed lines).
What is immediately apparant from Fig.~\ref{fig - MRF summary} is that the actual values are several orders of magnitude smaller than the optimized values. The enormity of this gap is largely due to the assumption of minimal damping. Inclusion of inhomogeneous broadening results in limits represented by the lower set of dashed lines in Fig.~\ref{fig - MRF summary}. This reduces the limits by approximately four orders of magnitude, making the gap about a factor of 100.  Conversely, this suggests that one avenue for substantially enhancing the MRF of real materials is to decrease inhomogenous broadening.

The gap seen in Fig.~\ref{fig - MRF summary} is reminiscent of the corresponding gap for the first and second hyperpolarizabilities in nonlinear optics.\cite{kuzyk13.01} In the theory of fundamental limits of the hyperpolarizabilities, one defines the \textit{intrinsic} hyperpolarizabilities as
\begin{align}
\beta_\text{int.} = \frac{\beta}{\beta_\text{max}}\text{ and }\gamma_\text{int.} = \frac{\gamma}{\gamma_\text{max}},
\end{align}
where $\beta_\text{max}$ and $\gamma_\text{max}$ are the maximum possible hyperpolarizabilities. The intrinsic hyperpolarizabilities have been used to augment experimental studies of  the nonlinear optical response of quantum materials.\cite{zhou08.01, clays01.01, chen04.01,perez11.02} Furthermore, exploring the gap between the empirically measured $\beta_\text{int.}$ and the maximum value of $1$ has lead to experimental \cite{clays01.01, eisle05.01, sekar19.01, sekar18.01, luu05.01} and theoretical \cite{serki01.01, crowe18.01, lytel15.02, burke13.01, zhou07.02} work aimed to improving nonlinear materials. This motivates the definition of the scale-invariant \textit{intrinsic material response factor}
\begin{align}
\zeta_\text{int.} &= \frac{\zeta}{\zeta^\text{3L}_\text{opt.}},\label{eqn - intrinsic MRF definition}
\end{align}
where $\zeta^\text{3L}_\text{opt.}$ is the MRF optimized within the TLM, where the natural linewidth is used for the damping factor to get the absolute upper bound.  We propose the intrinsic material response factor as a figure of merit for materials involved in radiative heat transfer.  This figure of merit can be used by researchers as a goal post for developing new molecules that can be used in the fabrication of nano-structures, and it provides a scale-independent method for comparing materials.  

We can interpret the intrinsic MRF in terms of spatial scaling in light of the dependence of the limits on $E_{10}$. The energy difference $E_{10}$ defines a characteristic length which measures the spatial extent of the electronic ground state wave-function.\cite{kuzyk16.01} Thus, instead of thinking of the optimized MRF as a limit for a given energy spectrum, it can be thought of as the best case scenario for the scaling of the MRF with respect to the size of the molecule. Molecules with a large intrinsic MRF are the elements that will scale most favorably as the size is increased. Thus, optimizing molecules using the intrinsic MRF is a two-step process. First, the intrinsic MRF is itself optimized. The best molecular units found can then be ``scaled up" to larger unit cells that might yield exceptionally large MRF. This method has been successfully utilized in nonlinear optics.\cite{perez16.01,perez16.02,slepk04.01}

\textit{Conclusions} -- In summary, we have demonstrated that the figure of merit arising from the fundamental limits to radiative heat transfer -- namely, the material response factor -- is itself amenable to constrained optimization, and that the constraints ultimately derive from the algebraic structure of quantum mechanics.  This limit depends only on the optical properties of the molecular substituents and the molecular number density. The limits found when compared with common materials are orders of magnitude higher, suggesting that molecular synthesis is a promising avenue for fabricating better materials for RHT. The prolific work that has previously been carried out in maximizing the linear and nonlinear response of molecules can be leveraged in the effort to maximize radiative heat transfer between amorphous materials. We find that the three level model provides a relatively simple theory that defines key desirable features in the energy spectrum and transition moments, which can be used to guide materials development and gain insight into the origins of the maximal material properties.

\textit{Acknowledgments} -- The authors would like to thank Dr. Sean Mossman for insightful discussions on the subject.

\bibliography{\bibs}

\end{document}